\documentclass[aps,prl,groupedaddress,floats,twocolumn]{revtex4-1}
\usepackage{amsmath,amsfonts,graphicx,mathrsfs,epstopdf, latexsym,bm,float}

\usepackage[colorlinks,allcolors=blue]{hyperref}

\begin{document}

\title{Enhanced Self-seeding with Ultra-short Electron Beams}

\author{Erik Hemsing}
\email{ehemsing@slac.stanford.edu}
\affiliation{SLAC National Accelerator Laboratory, Menlo Park, California 94025, USA}
\author{Aliaksei Halavanau}
\email{aliaksei@slac.stanford.edu}
\affiliation{SLAC National Accelerator Laboratory, Menlo Park, California 94025, USA}
\author{Zhen Zhang}
\affiliation{SLAC National Accelerator Laboratory, Menlo Park, California 94025, USA}

\date{\today}

\begin{abstract}
We describe a new method to produce intensity stable, highly coherent, narrow-band x-ray pulses in self-seeded free electron (FEL) lasers. The approach uses an ultra-short electron beam to generate a single spike FEL pulse with a wide coherent bandwidth. The self-seeding monochromator then notches out a narrow spectral region of this pulse to be amplified by a long portion of electron beam to full saturation. In contrast to typical self-seeding where monochromatization of noisy SASE pulses leads to either large intensity fluctuations or multiple frequencies, we show that this method produces a stable, coherent FEL output pulse with statistical properties similar to a fully coherent optical laser.
\end{abstract}

\maketitle
Self-seeded x-ray free electron lasers (FELs) use relativistic electron beams to produce intense, narrow-band x-ray pulses for a wide array of high resolution science applications. They work by effectively splitting a single pass, high-gain FEL into two sections. In the first section (i.e., the SASE FEL), the electron beam produces x-rays by self-amplified spontaneous emission (SASE). Originating from shot noise, the SASE pulse is stochastic in nature and has the features of chaotic polarized radiation~\cite{saldin1998statistical}. As such, it contains many uncorrelated temporal and frequency spikes. The SASE spectrum is then frequency-filtered with a monochromator (diffraction grating or crystal), which isolates a narrow region of SASE frequencies for amplification~\cite{feldhaus1997possible,geloni2011novel}. In the second section (the seeded FEL), the filtered light (i.e., the seed) is placed back on the electron beam and amplified to saturation. This technique has been used increase the coherence of x-ray FELs and to produce pulses with relative bandwidths $\sim 10^{-4}$ at both hard and soft x-ray wavelengths~\cite{ratner2015experimental,amann2012demonstration}.


Generally, the first-stage SASE pulse is produced in the linear regime by an electron beam (e-beam) that is much longer than the FEL cooperation time, $\tau_c=1/2\sqrt{3}\rho \omega_0$, where $\omega_0$ is the radiation frequency and $\rho$ is the FEL parameter~\cite{bonifacio1984collective}. The cooperation time is the slippage accrued over an exponential gain length and sets the approximate temporal scale over which the amplified radiation is coherent. Long flat e-beams with duration $T\gg 2\pi \tau_c$ contain a large number of coherent temporal spikes. Accordingly, the SASE spectrum, which spans a bandwidth $\sigma_A\approx\rho\omega_0=1/2\sqrt{3}\tau_c$, contains a large number of frequency spikes, each with average width $\Delta\omega_c=2\pi/T$. 
The average number of spikes that then pass through the self-seeding filter depends on the monochromator (mono) linewidth $\sigma_m$. 
If $\sigma_m\ll\Delta\omega_c$ 
then, on average, the seed consists of a single coherent spike, but exhibits large intensity fluctuations. If the filter bandwidth is much larger than a single frequency spike, then the seed contains multiple spikes, but the integrated  fluctuations are reduced. 
Thus, with SASE from long beams there is a fundamental trade-off between number of spikes (coherence) and fluctuations (stability). For the soft x-ray (SXR) self-seeding system at LCLS~\cite{ratner2015experimental}, the effective mono bandwidth ($\sigma_m\approx 85$~meV rms at 1~keV photon energies) 
passes $M=4-5$ coherent modes with the nominal $T\geq 50$~fs e-beam, 
and exhibits the associated $\sigma_\mathcal{E}=1/\sqrt{M}$ level of relative pulse energy fluctuations during exponential growth in the seeded stage~\cite{zhangstatexp2020}. While saturation effects can eventually reduce the fluctuations in the amplified seed, it lacks full temporal coherence. This precludes maximal energy extraction with strong downstream undulator tapering~\cite{PhysRevAccelBeams.19.020705}, and the buildup of SASE in the regions between coherent spikes generates an undesirable spectral pedestal~\cite{ratner2015experimental,hemsing2019statFEL}.

 \begin{figure}
   \includegraphics*[width=85 mm]{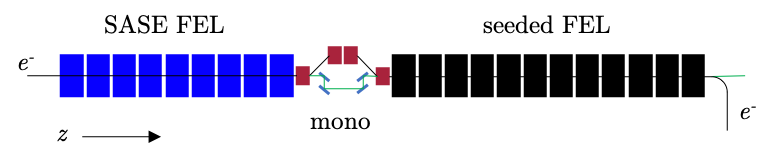}
   \includegraphics*[width=65 mm]{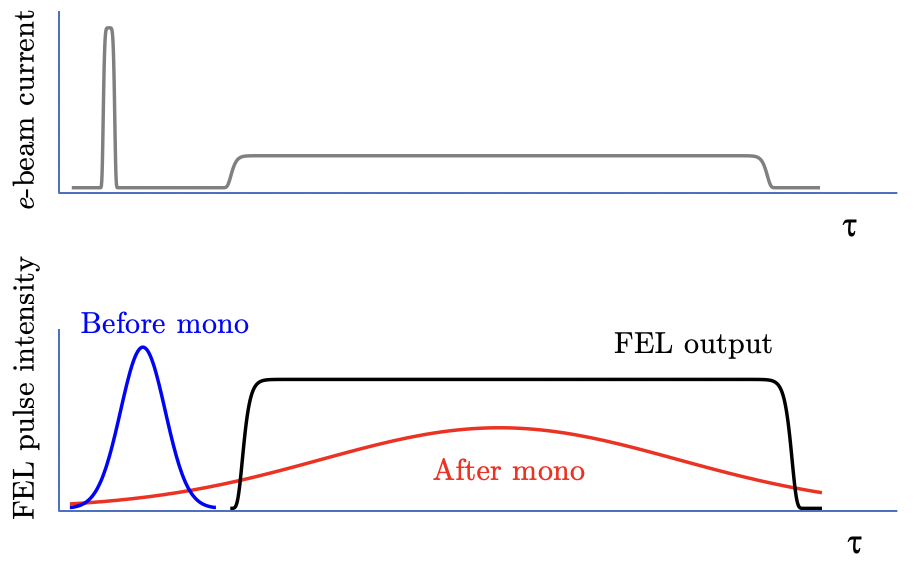}
   \caption{Illustration of the proposed scheme. Top: Layout of the SXR self-seeding beamline (not to scale). Below: A high-current, short pulse portion of  e-beam lases to saturation in the SASE section to produce a single spike pulse (blue). After the narrow-band monochromator (mono), the spike is stretched (red) to overlap a longer flat-current section of e-beam, and then amplified to saturation, producing a stable single mode long pulse in the FEL output (black). The long portion of the beam does not lase strongly in the SASE section due to low current and/or strong undulator tapering, but does lase strongly in the seeded section.}
   \label{scheme}
\end{figure}

 \begin{figure}
   \includegraphics*[width=.99\linewidth]{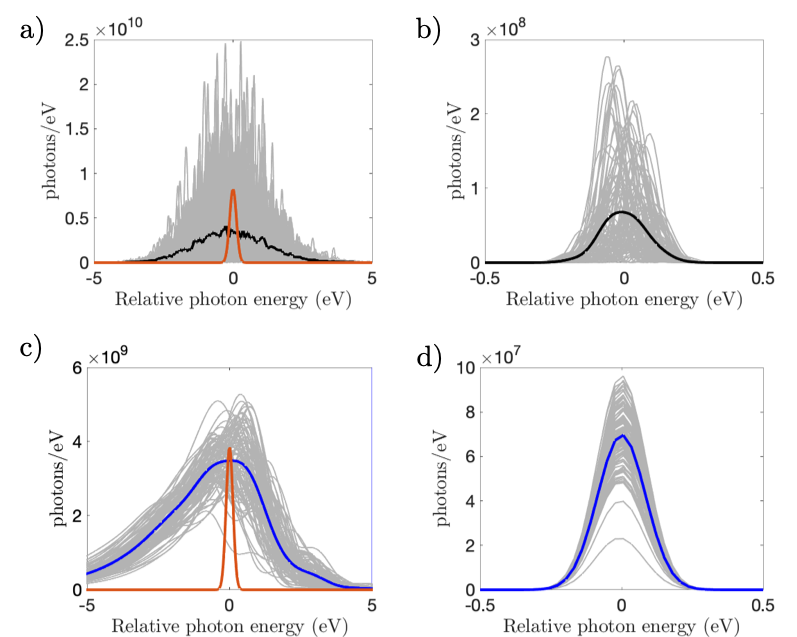}
   \caption{Comparison of spectra when seeding with a long SASE pulse (black) versus a saturated short SASE pulse (blue) at 1~keV photon energy modeled with 100 {\sc ginger} simulations. With a long e-beam $T\gg 2\pi \tau_c$ (a), the incident SASE frequencies on the mono filter (red) fluctuate stochastically, resulting in a fluctuating seed (b). With a short e-beam $T= 2\pi \tau_c$ (c), the incident single spike spectrum is much more stable, as is the single-mode seed (d).}
   \label{SASEvsSPS}
\end{figure}

Here we propose a simple new self-seeding scheme that reduces the level of fluctuations to the few percent level while also maintaining a temporally coherent FEL output. 
In this arrangement the FEL statistically behaves less like a chaotic source and more like an ideal laser~\cite{Glauber1963} or an externally seeded FEL~\cite{Gorobtsov2018fermistat}. Inspired by the proposal to produce single spike FEL pulses in Ref.~\cite{ROSENZWEIG2008short} and recent work on the generation of stable sub-fs pulses at LCLS~\cite{HuangSingleSpikeHXR2017,duris2019xleap}, this scheme uses an ultra short, high current portion of electron beam with duration
\begin{equation}\label{eq:T_coh}
    T\approx 2\pi \tau_c=\pi/\sqrt{3}\rho\omega_0
\end{equation}
to produce a saturated, single spike ($M\approx 1$) SASE x-ray FEL pulse incident on the self-seeding mono. The mono then selects a narrow section of the coherent spectrum to act as a seed, essentially acting as a pulse stretcher of the saturated single spike SASE pulse. The concept is illustrated in FIG.~\ref{scheme}. The self-seeding chicane delay is adjusted to place the seed on a different, longer portion of e-beam with lower current that amplifies the narrow-band seed to saturation. This scheme requires no additional x-ray optics, only tailoring of the incoming e-beam in a manner similar to previous efforts, including chirp/taper approaches~\cite{MacArthur2019prl,Zhang2019prab}.

In this technique it is important that lasing of the long portion of the e-beam in the SASE section is weak so that it remains unspoiled for seeding. This may be accomplished simply by sufficiently low current, or by active suppression, e.g., by laser heater shaping~\cite{MarinelliLaserHeater2016}, with the fresh-slice technique~\cite{lutman2016fresh}, or even with a strong undulator taper~\cite{PhysRevSTAB.8.040702}. Likewise, it is desirable to prevent further FEL emission in the seeded stage by the spent high current spike. In some self-seeding designs, dispersion in the chicane, used to compensate the optical delay, can decompress the current spike by exploiting the large energy spread at saturation.

The distinction between this approach and typical self-seeding is illustrated in FIG.~\ref{SASEvsSPS}. We find that this technique, which is a version of fresh bunch self-seeding~\cite{cemmafreshbunchSS2017}, has several advantages.
First, the short e-beam produces a single spectral spike pulse that fully covers the seeded FEL bandwidth, so there is always a single spectral mode with significant power within the mono bandwidth. Second, the high current spike reaches saturation at the mono, which significantly reduces intensity fluctuations of the seed compared to normal self-seeding, where the signal is filtered early in the linear regime. Third, because the short pulse saturates, the peak seed power can be orders of magnitude higher at the start of the seeded section than in normal self-seeding, but with only modest pulse energy. The higher power seed helps stabilize the final FEL output by vastly exceeding the shot noise power (here by $10^3$), while the x-ray energy deposition on the self-seeding optics is kept small to prevent damage. Fourth, seeding with a high power, stable single mode then enables reliable downstream amplification and strong tapering to maximize FEL output within the narrow seeded bandwidth, and with minimal spectral SASE pedestal. Finally, this scheme enables some unique coherent tailoring of the FEL seed. For example, with soft x-rays, different grating surface designs would allow near-transform limited tuning of the FEL output by varying $\sigma_m$ without changing the number of seed modes, or even seeding multiple phase-stable colors. 

This concept builds on the statistical properties of ultra-short SASE FEL pulses near saturation, which have been studied previously. 
In Ref.~\cite{saldin2003shortpulse}, numerical studies indicated that relative pulse energy fluctuations in the saturation (nonlinear) regime are reduced to the $\sigma_\mathcal{E}=10-20\%$ level when $T$ shrinks to become comparable with $2\pi\tau_c$. These results are in agreement with experimental measurements~\cite{AyvazyanStatFLASH2003}, and are attributed to the strong slippage effects. 
In the linear regime, slippage smooths the initial shot noise, but there are still fluctuations in the onset of lasing.
Eventually, however, the power and e-beam bunching saturate and the coherent pulse slips out front of the e-beam, significantly stabilizing the output a few gain lengths beyond saturation. 
We find that, after spectral filtering through the narrow-band mono, these pulses serve as stable and coherent seeds for subsequent amplification. 


\begin{table}
\centering
  \begin{tabular}{l c c}\hline
\hline
Parameter &	Value &	Units   \\
\hline
Beam energy, $E$ & 4 & GeV \\
Undulator strength (rms) K & 2.29226 & - \\
Undulator period, $\lambda_u$ & 3.9 & cm \\
Undulator section length, $L_u$ & 3.4 & m \\
Undulator drift length, $L_{ud}$ & 1.0 & m \\
Fundamental energy, $\hbar\omega_0$ & 620 & eV \\
Monochromator bandwidth (rms), $\sigma_m$ & 85 & meV \\
\hline
\hline
\end{tabular}
\caption{\label{table-sxr}LCLS-II SXR beamline parameters}
\end{table}

 \begin{figure}
    \centering
 \includegraphics[width=0.99\linewidth]{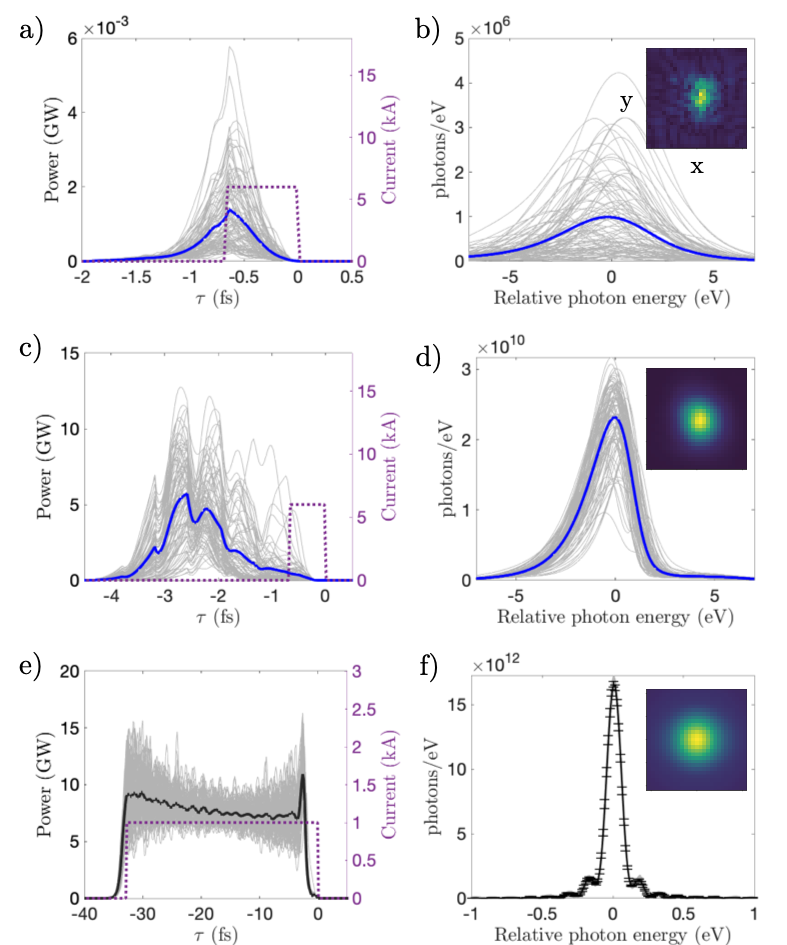}
    \caption{Radiation temporal profiles and spectra of the short SASE pulse (blue) at $z=10$~m (a,b) and $z=40$~m (c,d), and of the amplified seed (black) at the $z=80$~m saturation point (e,f) for 100 {\sc ginger} simulations of lasing at 620~eV. The flattop electron beam current profiles are shown with dashed lines, with the head to the left. Error bars on the seeded FEL output spectrum are $\pm1$ rms fluctuations at each frequency. The inset shows the single shot transverse radiation mode intensity over a 100~$\mu$m window from {\sc Genesis}.}
    \label{fig:field_t}
\end{figure}

To illustrate, we consider the case of the LCLS-II SXR beamline at SLAC (see TABLE.~\ref{table-sxr}). We modeled the system in the conventional FEL codes {\sc genesis}~\cite{Genesis} and {\sc ginger}~\cite{ginger}, with both showing nearly identical results. Lasing is simulated at a $\hbar\omega_0=620$~eV fundamental photon energy ($\lambda = 2$ nm) 
in both the SASE and seeded sections. First, the short pulse SASE section is simulated over multiple runs to generate a statistical dataset for analysis. The output fields are then monochromatized using a $2\%$ efficiency, $\sigma_m$=85~meV rms width Gaussian filter (identical to the LCLS design) positioned at the average spectral peak to extract the seed fields. These 3D fields are then used as inputs to be amplified by the longer portion of e-beam in the seeded simulations. The current profile of the short e-beam is a $T=2\pi\tau_c=0.67$~fs duration flat-top with 6~kA peak current and 3~MeV energy spread ($\rho=2.7\times10^{-3}$, $\tau_c=0.11$~fs). This corresponds to a compressed portion of the long e-beam, which is a 30~fs flat-top with 1 kA peak current and 0.5 MeV energy spread ($\rho=1.5\times10^{-3}$). This two e-beam arrangement (shown in FIG.~\ref{scheme}) is similar to recent experiments that exploit short range wakefields in a single e-beam to produce sub-fs pulses~\cite{duris2019xleap,MacArthur2019prl}, or may be produced by tailoring double-bunch configurations~\cite{PhysRevAccelBeams.20.030701,Halavanau:co5121}. The beta function of both beams is 12~m, and a 0.35~$\mu$m normalized transverse emittance is assumed, though at these example SXR wavelengths the FEL performance is relatively insensitive to the emittance. We note that 
 the short SASE pulse is fairly insensitive to the precise shape the ultra-short beam current profile due to the strong slippage effects.

 \begin{figure}
    \centering
\includegraphics[width=0.99\linewidth]{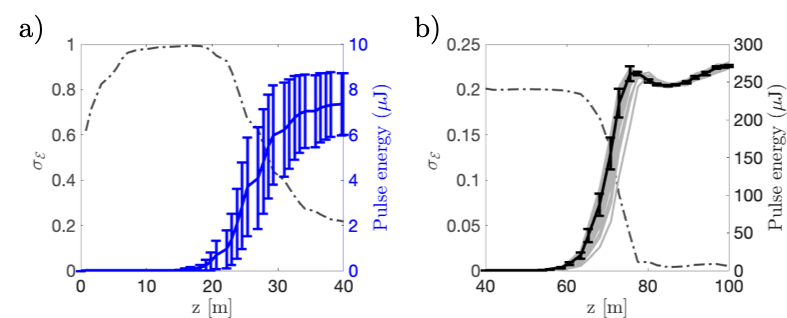}
\caption{Statistical fluctuations and gain curve for short SASE pulse (a) and seeded pulse (b) simulated for LCLS-II with {\sc ginger}. After the seed is monochromatized at the $z=40$~m point, it has $\sigma_\mathcal{E}=20\%$ fluctuations and 1~MW power. After amplification to 8~GW, the fluctuations are at the $1\%$ level.}
    \label{fig:field_xy2}
\end{figure}

The temporal and spectral profiles of the short pulse and of the amplified seed pulse are shown in FIG.~\ref{fig:field_t}. Shot noise is included in all simulations. Early in the linear regime, the short pulse exhibits a well-known ``shark fin" temporal profile (FIG.~\ref{fig:field_t}a) from the flattop short e-beam \cite{BonifacioTimeDepend1988,CaiTimeDepend1990}. Due to slippage, the pulse evolves into a 2~fs long, 6~GW ($7~\mu$J) Gaussian-like pulse near saturation (FIG.~\ref{fig:field_t}c). It has small temporal modulations from the 1~m drifts between 3.4~m long undulator modules. By saturation, the radiation mode is fully formed and has near full transverse coherence (FIG.~\ref{fig:field_t}d inset)~\cite{PhysRevA.35.3406,SALDIN2000185}. At $z=40$~m the short pulse is frequency-filtered through the mono, which stretches it to a Gaussian temporal pulse with 9~fs fwhm duration and 1~MW average peak power ($0.01~\mu$J pulse energy). It is then amplified by the 30~fs long e-beam, for which the effective shot noise power is about 0.5~kW. By saturation in the seeded stage, the 8~GW ($250~\mu$J) radiation pulse displays the flattop temporal profile of the long e-beam (FIG.~\ref{fig:field_t}e), and has an associated intensity-stable, highly coherent narrow-band spectrum with 120 meV FWHM bandwidth (FIG.~\ref{fig:field_t}f).

The evolution of the pulse energy and of the fluctuations in each stage is shown in FIG.~\ref{fig:field_xy2}. The short pulse shows strong fluctuations in the linear regime that drastically reduce around $z=30$~m where it starts to saturate (FIG.~\ref{fig:field_xy2}a). Deeper in saturation, the short pulse is most stable ($\sigma_\mathcal{E} \approx 20 \%$).
 In the seeded FEL, the statistical intensity fluctuations are fixed by the input seed statistics through the linear regime, and then drop significantly as the system saturates. By the end of the LCLS-II SXR undulator ($z>70$~m), the fluctuations are $\sigma_\mathcal{E} \approx 1\%$. This performance stability resembles a conventional optical laser~\cite{Goodman}. 

\begin{figure}
    \centering
    \includegraphics[width=0.99\linewidth]{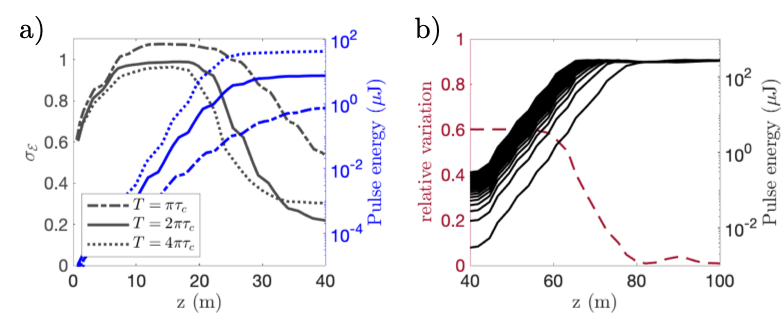}
    \caption{(a) Total average pulse energy in the SASE section and relative fluctuations through a narrow-band filter for different durations $T$ of the short e-beam. Results are from 500 {\sc ginger} simulations for each $T$. (b) FEL pulse energy in the seeded section for different input seed pulse energies. The rms variation in FEL pulse energy across these values is also shown. It drops to $1\%$ near the $z=80$~m point, indicating little sensitivity to the input seed power level.}
    \label{fig:fluct}
\end{figure}

An important question is how sensitive the FEL output is to the length of the ultra-short e-beam. As shown in FIG.~\ref{fig:fluct}a, for a fixed 6~kA current, we find that a change in the e-beam length over the range $2\pi\tau_c\leq T\leq 4\pi\tau_c$ has little impact on the relative fluctuations through the narrow-band mono, similar to Ref.~\cite{saldin2003shortpulse}. Beams much shorter than $2\pi\tau_c$ do not reach saturation at the mono position and thus have larger fluctuations. Beams longer than $4\pi\tau_c$ begin to develop multiple frequency spikes which also increases fluctuations within the narrow line width. 

Another issue is the robustness of the seeded FEL to variations in seed power. In practice, if the short e-beam duration or current fluctuates (e.g., from compression jitter), it changes the spectral brightness (photons/eV) at the mono, and therefore the pulse power in the seed. For example, in the LCLS-II case, the filtered seed power for the pulse from the $2\pi\tau_c$ beam is 1~MW, while for the $4\pi\tau_c$ beam it is 4.5~MW. This issue is explored in FIG.~\ref{fig:fluct}b, where we show results from a scan over the seed pulse power from 0.1-10~MW (0.002-0.2$~\mu$J) in steps of 0.4~MW. During exponential gain the seeded FEL power depends on the input seed power. Shortly after saturation at the $z=80$~m mark, however, it is virtually independent of the seed power, as indicated by the $<3\%$ rms variation in energy over the full range of input powers. The spectra are all virtually identical. This indicates that, when seeded with the single spike, high power pulse from the short beam, the saturated seeded FEL is highly insensitive to fluctuations in seed power, as expected~\cite{PhysRevLett.57.1871}.

\begin{figure}
    \centering
     \includegraphics[width=0.99\linewidth]{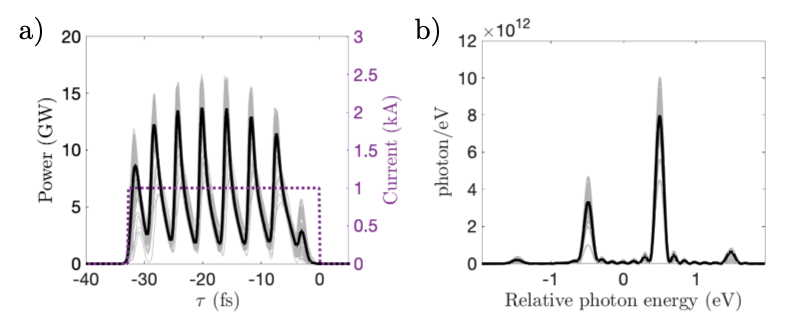}
    \caption{Temporal (a) and spectral (b) amplified two-color seed pulse profiles at saturation ($z=80$~m). Results are from 100 {\sc ginger} LCLS-II simulations using two colors filtered from the short pulse spectrum centered at 620~eV.}
    \label{fig:twocolor}
\end{figure}

The broad coherent bandwidth incident on the mono allows the possibility to tailor the coherent properties of the FEL, similar to techniques pioneered in external seeding (e.g.,~\cite{Gauthierphaselocked2016}). With customized monochromators, one could envision self-seeding with tunable coherent bandwidths or multiple well-defined and phase-stable frequencies. For example, at SXR wavelengths a dual-color grating (enabled by two superimposed or alternating line-densities) could select two colors within FEL bandwidth, similar to two-color seeding explored at hard x-rays~\cite{lutman2014demonstration}. Shown in FIG.~\ref{fig:twocolor} are the results of simulations where the 620~eV short pulse spectrum in FIG.~\ref{fig:field_t}d is filtered by two $\sigma_m=85$~meV Gaussian apertures separated by 1~eV. The total 0.5~MW input power is split between the colors, which produce a beat modulation on the temporal profile in the seeded section. This modulation persists over multiple shots, indicating high phase stability between the two colors. Similar to a single-color, the total energy fluctuations in the amplified two-colors drop to $\sigma_\mathcal{E}\leq5\%$ at saturation. Further into saturation, additional sideband frequencies also appear at $1$~eV intervals out to the edges of the FEL bandwidth, but do not strongly impact the total statistics. Though not observed in this example, we note that in regions where the seeding is weak, such as near the e-beam head or tail or within the temporal beat, SASE can develop and produce a small spectral pedestal, especially near saturation.


In summary, we find that the proposed enhanced self-seeding approach may be used to overcome fundamental limitations in conventional self-seeding.  By spectrally filtering pulses from an ultra short e-beam at saturation, this enables highly stable, fully coherent, and customizable x-rays from modern FELs.
The authors acknowledge helpful assistance from W.~M.~Fawley and G.~Marcus on simulations, and J.~Hastings, S.~Serkez, and G.~Wilcox on potential two-color grating designs. This work was supported by U.S. Department of Energy Contract No. DE-AC02-76SF00515 and award no. 2017-SLAC-100382.


\bibliography{StatisticalFEL}

\end{document}